\documentstyle[prl,aps,psfig]{revtex}
\newcommand{\beq}{\begin{equation}}
\newcommand{\eeq}{\end{equation}}
\newcommand{\beqa}{\begin{eqnarray}}
\newcommand{\eeqa}{\end{eqnarray}}
\newcommand{\ba}{\begin{array}}
\newcommand{\ea}{\end{array}}

\begin{document}
\draft

\twocolumn[\hsize\textwidth\columnwidth\hsize\csname
@twocolumnfalse\endcsname

\widetext 
\title{Dimensional Reduction in Bose-Condensed Alkali-Metal Vapors} 
\author{L. Salasnich$^{1}$, A. Parola$^{2}$ and L. Reatto$^{1}$} 
\address{$^{1}$Istituto Nazionale per la Fisica della Materia, 
Unit\`a di Milano, 
Dipartimento di Fisica, Universit\`a di Milano, \\
Via Celoria 16, 20133 Milano, Italy\\
$^{2}$Istituto Nazionale per la Fisica della Materia, 
Unit\`a di Como, 
Dipartimento di Scienze Fisiche, Universit\`a dell'Insubria, \\
Via Valeggio 11, 23100 Como, Italy}

\maketitle

\begin{abstract} 
We investigate the effects of dimensional reduction 
in atomic Bose-Einstein condensates (BECs) 
induced by a strong harmonic confinement in the cylindric 
radial direction or in the cylindric axial direction. 
The former case corresponds to a transition from 3D to 1D 
in cigar-shaped BECs, while the latter case 
corresponds to a transition from 3D to 2D in disc-shaped 
BECs. We analyze the first sound velocity 
in axially-homogeneous cigar-shaped 
BECs and in radially-homogeneous disc-shaped BECs. 
We consider also the dimensional reduction 
in a BEC confined by a harmonic potential 
both in the radial direction and in the axial direction. 
By using a variational approach, we calculate 
monopole and quadrupole collective oscillations of the 
BEC. We find that 
the frequencies of these collective oscillations are related 
to the dimensionality and to the repulsive or attractive 
inter-atomic interaction. 
\end{abstract}


]

\narrowtext
 
In the experiments with Bose-Einstein condensates (BECs) 
the atomic cloud is confined by a magnetic or magneto-optical 
trap. In the simplest case the confining field acts as an 
anisotropic harmonic potential. 
Recent experiments \cite{1,2} have shown that it 
is possible to change the frequencies of the harmonic 
confinement inducing a dimensional reduction of the system: 
from 3D to 1D and from 3D to 2D. 
Effects of reduced dimension in BECs have been 
theoretically investigated 
by many authors \cite{3,4,5,6}, 
but less attention has been paid to the transition 
from a dimensional regime to another. 
\par 
In this brief report we study in detail 
the dimensional reduction in BECs 
by using nonpolynomial Schr\"odinger equations, which are 
derived from the 3D Gross-Pitaevskii equation (GPE) \cite{7}. 
In this way we calculate chemical potential, 
elementary excitations and first sound velocity for 
cigar-shaped and disc-shaped BECs. Moreover, 
we apply a variational technique \cite{8} to obtain 
monopole and quadrupole collective excitations of the BEC 
with both positive and negative scattering length $a_s$. 
In particular, we analyze the effects of anisotropy and 
inter-atomic strength on these collective frequencies. 
\par
{\it From 3D to 1D}. -- 
We have recently shown that the dynamics of a cigar-shaped 
BEC of $N$ atoms, confined by a harmonic potential 
of frequency $\omega_r$ in the cylindric radial 
direction $(x,y)$ and by a generic potential $V(z)$ 
in the cylindric axial direction $z$, is well described 
by an effective 1D nonpolynomial Schr\"odinger equation (1D NPSE) 
for the axial wave function $f(z,t)$ of the BEC \cite{7}. 
Within the 1D NPSE approach, the BEC is approximated 
by a Gaussian in the cylindric radial direction. The Gaussian 
width $\sigma$ is related to $f$ by 
$\sigma=(1+gN|f|^2)^{1/4}$, 
where $g=2 a_s/a_r$ is the effective interaction strength 
with $a_r=\sqrt{\hbar/(m\omega_r)}$ the characteristic 
harmonic length in the radial direction \cite{7}.
Note that we write length is in units $a_r$ 
and energy in units $\hbar \omega_r$. 
\par
In the limit of weak radial confinement, 
which corresponds to the 3D regime, 
the axial density $N|f|^2$ satisfies the condition 
$gN|f|^2 >> 1$ and the 1D NPSE reduces to a 1D nonlinear 
Schr\"odinger equation with quadratic nonlinearity. In the opposite 
limit $gN|f|^2 << 1$ of strong radial confinement, 
which corresponds to a 1D regime, 
the 1D NPSE becomes a 1D nonlinear Schr\"odinger equation 
with cubic nonlinearity, also called 1D GPE \cite{4}. 
It is important to remember that the 1D NPSE is derived from 
the 3D GPE, but the 3D GPE (and also 1D NPSE) 
is no more valid in the limit of strong diluteness: in this case 
the atomic cloud becomes a Tonks gas, i.e. a gas of impenetrable 
Bosons \cite{9}. In our units the system is 
far from the Tonks regime if the condition $gN|f|^2 \gg g^2$ 
is satisfied [10]. 
\par 
In previous papers \cite{7} we have verified that the 
1D NPSE is able to capture the full 3D regime, 
the full 1D regime (apart the Tonks gas) and the region in between. 
Here we show that an analytical expression for the shape 
of the axial density can be obtained by neglecting the kinetic term, 
the so-called Thomas-Fermi (TF) approximation, in the 1D NPSE. 
Then, setting $f(z,t)= e^{-i\mu t} f_0(z)$ one finds 
for the axial density profile $\rho_1(z)=N|f_0(z)|^2$ 
the following expression 
\beq 
\rho_1(z) = {2 \over 9 g} 
\left[ \mu(z)^2 - 3 + \mu(z)\sqrt{\mu(z)^2+3} \right] \; ,  
\eeq  
where $\mu(z) = \mu - V(z)$ with $\mu$ the chemical potential. 
This density profile reduces 
to $\rho_1(z)=(2\mu(z))^2/(9g)$ when $g\rho_1 >> 1$, 
that is very close (with the factor $4/9$ 
instead of $1/2$) to the axial density profile of 
the TF 3D regime obtained from the 3D GPE. Eq. (1) reduces 
to $\rho_1(z)=(\mu(z)-1)/g$ when $g\rho_1 <<1$, 
that is exactly the density profile of the TF 1D regime obtained 
from 1D GPE.  
\par 
In the case $V(z)=0$, which corresponds to a homogeneous 
BEC in the $z$ direction with periodic boundary conditions, 
one gets the axial velocity of first sound as 
$ 
c_1 = ( \rho_1 {\partial \mu\over \partial \rho_1} )^{1/2}
$ 
form which one finds 
\beq 
c_1= 
\sqrt{ 
{5\over 4} {g\rho_1\over \sqrt{1+ g\rho_1} } 
- {1\over 4} {g\rho_1+2g^2\rho_1^2\over (1+ g\rho_1)^{3/2} } 
} \; . 
\eeq 
In Figure 1  we plot $c_1$ as a function of $g\rho_1$. 
The figure shows that 
$c_1$ strongly depends on the dimensionality. In fact, 
our expression for the velocity of sound gives 
$c_1=\sqrt{3/4}(g\rho_1)^{1/4}$ in the 3D regime  
and it becomes $c_1 = \sqrt{g\rho_1}$ in the 1D regime. 
Note that our 1D value for $c_1$ is in full agreement 
with the result of Menotti and 
Stringari \cite{10}, while at higher density our result 
$c_1$ differs from the exact one obtained by 
the 3D GPE \cite{10}, only for the constant that is 
$\sqrt{2}/2$ instead of $3/4$. This small discrepancy 
(relative error less than 3\%) in the 3D regime 
is a consequence of using in the 1D NPSE 
a Gaussian radial wave function. 
Thus, our Eq. (2) for the sound velocity $c_1$ 
interpolates very well between the exact results 
of low density (1D regime) and high density (3D regime). 

\begin{figure}
\centerline{\psfig{file=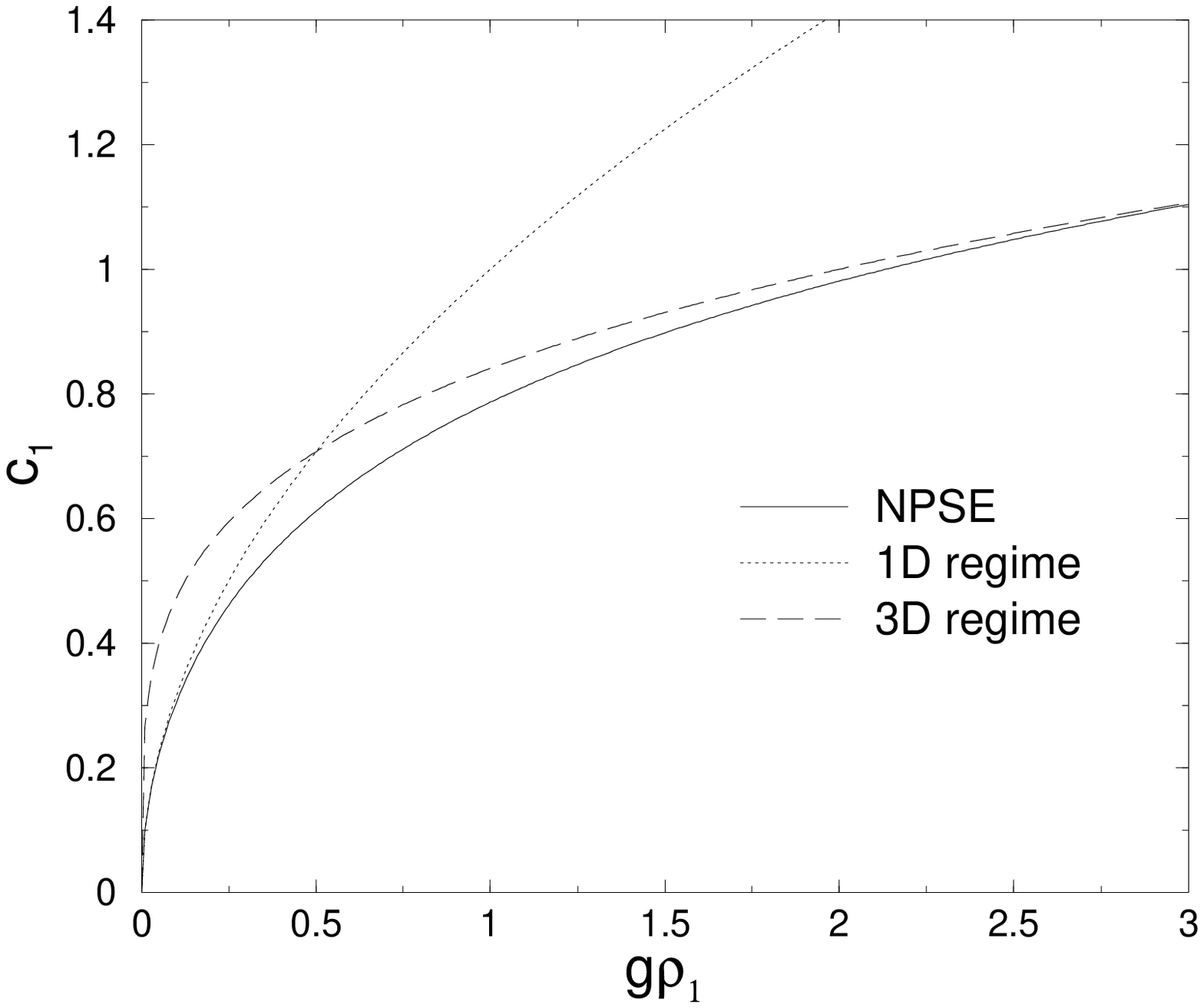,height=2.in}}
\small{FIG. 1: First sound axial velocity $c_1$ versus 
inter-atomic strength $g\rho_1$, where 
$g=2a_s/a_r$ with $a_s$ scattering length and 
$a_r$ radial harmonic length. 
$\rho_1$ is the axial density of the cigar-shaped 
axially-homogeneous BEC. 
Length in units $a_r$, density in units $1/a_r$ and 
time in units $1/\omega_r$. } 
\end{figure}  

In the case $V(z)=m\omega_z^2z^2/2$ 
the low-frequency elementary excitations, 
which are collective oscillation of the BEC, 
can be more easily calculated with a variational 
approach \cite{8} from the full 3D GPE instead of using the 
1D NPSE. 
The most natural choice is a Gaussian wave function 
with two variational parameters: 
the radial width $\sigma$ and the axial width $\eta$. 
Then the effective potential energy $U$ of the 
atomic cloud obtained from the 3D Gross-Pitaevskii 
energy functional is given by 
\beq 
U={1\over \sigma^2} + {1\over 2\eta^2} 
+ \sigma^2 + {1\over 2\lambda^2} \eta^2 
+ {N g\over \sqrt{2\pi}} {1\over \sigma^2 \eta} \; ,   
\eeq 
where $\lambda =\omega_r/\omega_z$ is the 
anisotropy parameter of the confining harmonic potential. 
With $\lambda >1$ the BEC is cigar-shaped, while 
for $\lambda <1$ the BEC is disc-shaped. 
The equilibrium point $(\sigma^*,\eta^*)$, corresponds 
to the minimum of the effective potential energy of the system. 
The low-frequency collective excitations of the BEC 
are the small oscillations around the equilibrium point. 
For $\lambda =1$ 
these two normal mode frequencies can be identified as 
the monopole frequency $\omega_M$ and the quadrupole frequency 
$\omega_Q$ \cite{8}. One finds that for $\lambda >>1$ the frequency 
$\omega_M$ is mainly transverse and the frequency 
$\omega_Q$ is mainly axial, while for $\lambda << 1$ the frequency 
$\omega_M$ is mainly axial and the frequency 
$\omega_Q$ is mainly transverse. 

\begin{figure}
\centerline{\psfig{file=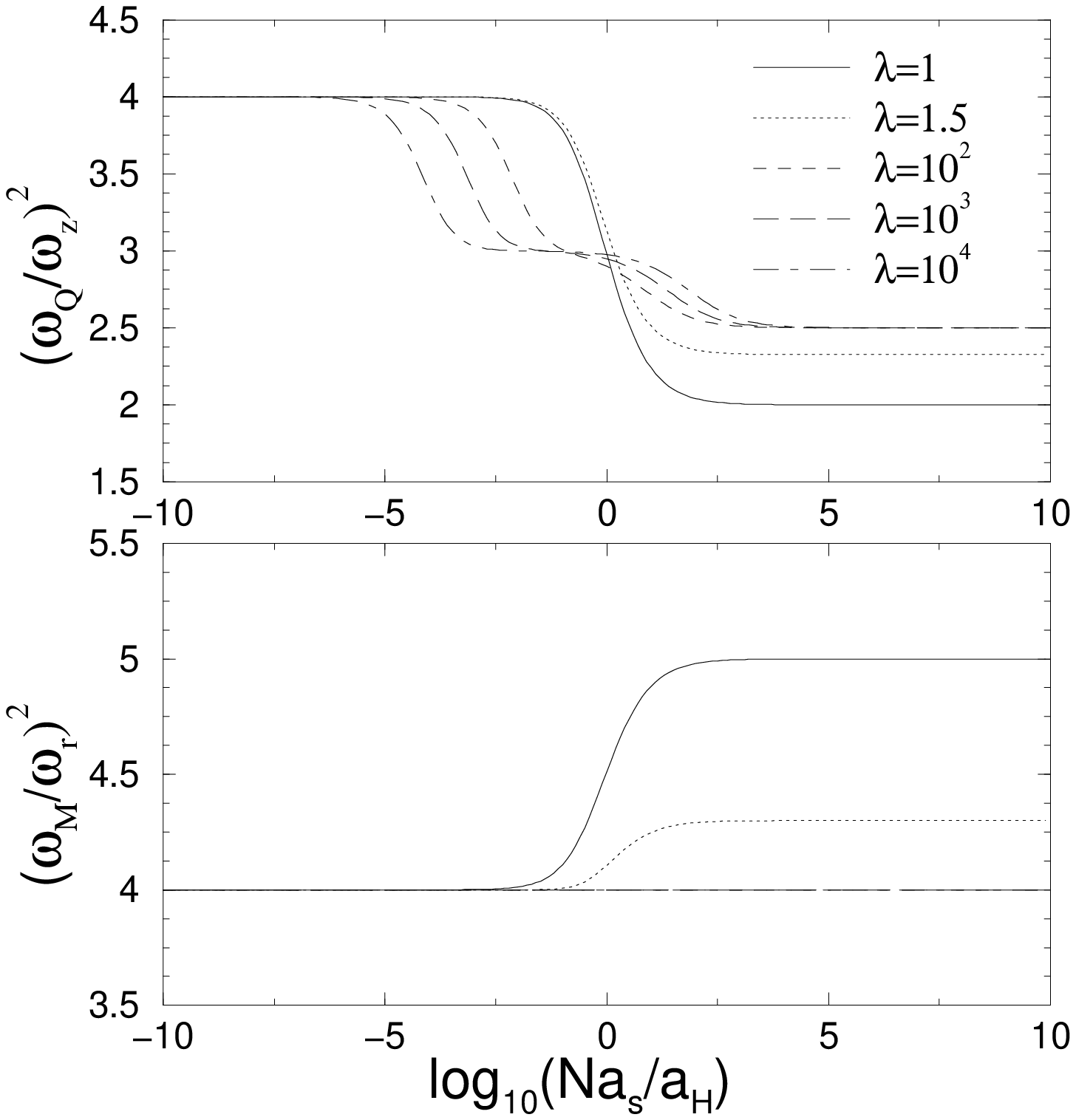,height=2.5in}}
\small{FIG. 2: Monopole frequency $\omega_M$ and quadrupole 
frequency $\omega_Q$ of collective oscillations 
for a cigar-shaped BEC. 
Case of positive scattering length $a_s$. 
$\lambda=\omega_r/\omega_z$ is the anisotropy of the 
harmonic trap. $a_H=\sqrt{\hbar/(m\omega_H)}$ with 
$\omega_H=(\omega_r^2\omega_z)^{1/3}$.} 
\end{figure} 

\par
In Figure 2 we plot $\omega_M$ and $\omega_Q$ 
as a function of the inter-atomic strength 
for different values of the anisotropy 
parameter $\lambda$. In the spherical case ($\lambda =1$) 
the monopole frequency $\omega_M$ ranges between 
the value $(\omega_M/\omega_r)^2=5$ of the 
3D regime of mean-field bosons (isotropic TF 3D) 
and the value $(\omega_M/\omega_r)^2=4$ of 
quasi-ideal bosons. 
By increasing $\lambda$ the frequency $\omega_M$ 
becomes a constant, given by $(\omega_M/\omega_r)^2=4$, 
independently on the interaction strength. 
\par 
The transition from 3D to 1D is even better 
shown by the quadrupole frequency $\omega_Q$ 
(see also [12]). In this case, for very large  
values of $\lambda$ the frequency 
$\omega_Q$ shows three plateaus: 
at $(\omega_Q/\omega_z)^2=5/2$ the BEC 
is in the 3D regime (anisotropic TF 3D), 
at $(\omega_Q/\omega_z)^2=3$ the BEC is 
in the regime of 1D confinement and mean-field bosons  
(TF 1D), and at $(\omega_Q/\omega_z)^2=4$ the BEC is 
in the regime of 1D confinement and quasi-ideal bosons 
(free bosons 1D). 
Thus Figure 2 shows that it is necessary 
a very strong anisotropy $\lambda$ 
to reach the regime of 1D confinement and mean-field bosons,  
and $\omega_Q$ is able to discriminate between 
this regime and that of 1D confinement and 
quasi-ideal bosons. Note that, as shown in [10], 
the Tonks regime is avoided if $N (a_s/a_H) \gg \lambda^{5/3} 
(a_s/a_H)^3$: the results of Fig. 2 are reliable for all values 
of horizontal axis under the condition 
$a_s/a_H \ll \lambda^{-5/9} 10^{-10/3}$. 
\par 
Our variational approach can be applied also to the case 
of an attractive BEC ($a_s<0$). In this case 
it exists a critical inter-atomic strength below which 
there is the collapse of the BEC \cite{11}, as 
shown in Figure 3. 

\begin{figure}
\centerline{\psfig{file=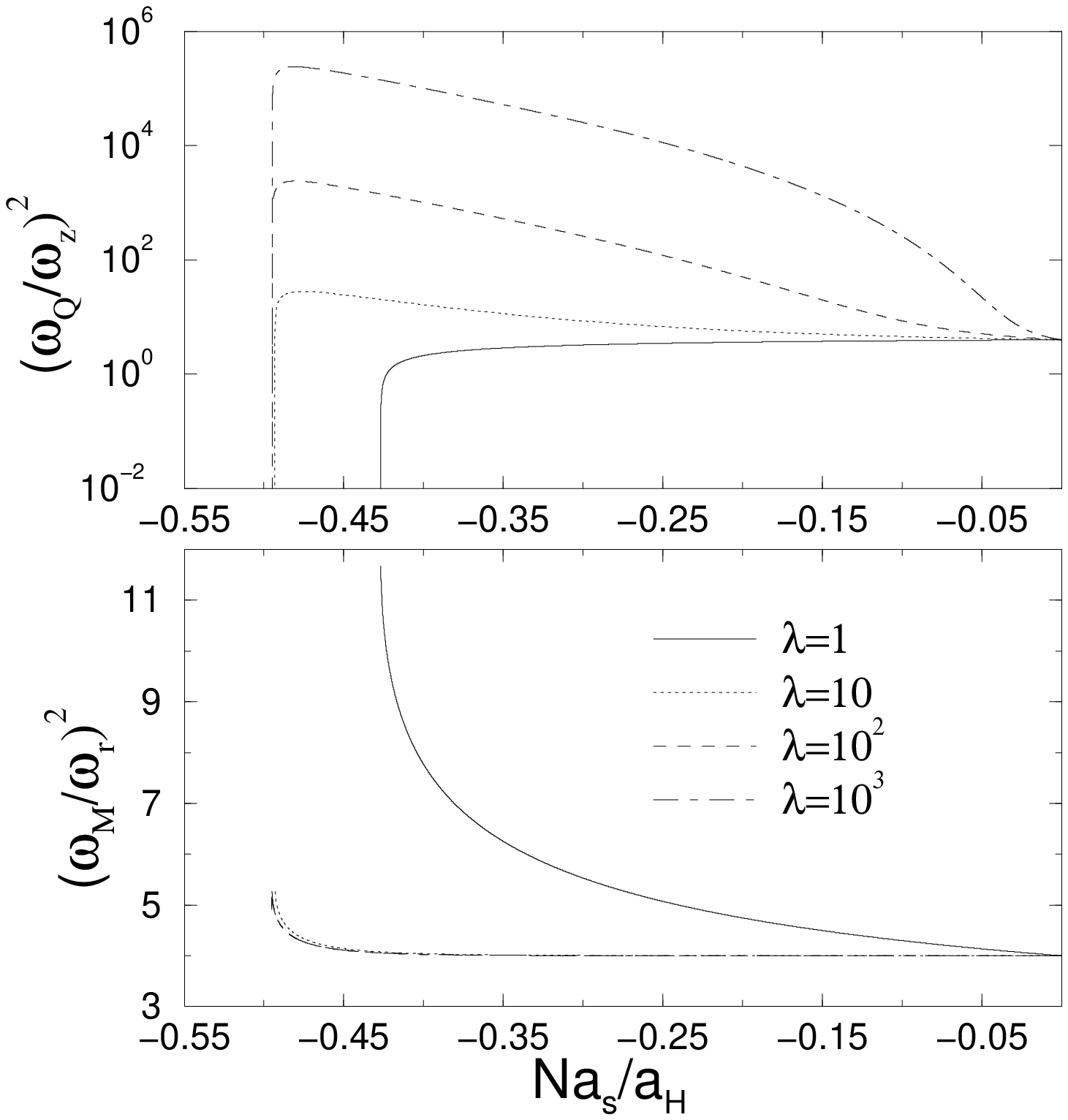,height=2.5in}}
\small{FIG. 3: Monopole frequency $\omega_M$ and quadrupole 
frequency $\omega_Q$ of collective oscillations 
for a cigar-shaped BEC. 
Case of negative scattering length $a_s$.} 
\end{figure}  

In the spherical case the monopole 
frequency $\omega_M$ grows by decreasing the interaction 
strength up to the collapse, but it is sufficient a small 
anisotropy to freeze $\omega_M$ to a constant value 
$(\omega_M/\omega_r)^2=4$. Instead 
the quadrupole frequency $\omega_Q$ strongly 
increases with $\lambda$ but always 
goes to zero at the collapse. 

{\it From 3D to 2D}. -- 
The dynamics of a disc-shaped BEC of $N$ atoms,  
confined by a harmonic potential 
of frequency $\omega_z$ in the cylindric axial 
direction $z$ and by a generic potential $W(x,y)$ 
in the cylindric radial direction $(x,y)$, is well described 
by an effective 2D nonpolynomial Schr\"odinger equation 
(2D NPSE) for the radial wave function $\phi(x,y,t)$ 
of the BEC. Within the 2D NPSE approach, the BEC is approximated 
by a Gaussian of width $\eta$ in the cylindric axial direction, 
given by $\eta^4 - \gamma N |\phi|^2 \eta -1  = 0$, 
where $\gamma = 2\sqrt{2\pi} a_s/a_z$ is the effective interaction 
strength and $a_z=\sqrt{\hbar/(m\omega_z)}$ the characteristic 
harmonic length in the radial direction. 
Here length is in units $a_z$ and energy 
in units $\hbar \omega_z$. 
\par
Note that in the limit of weak axial confinement, 
which corresponds to the 3D regime, 
the radial density $N|\phi|^2$ satisfies the condition 
$\gamma N|\phi|^2 >> 1$ and the 2D NPSE reduces to a 2D nonlinear 
Schr\"odinger equation with quadratic nonlinearity. In the opposite 
limit $\gamma N|\phi|^2 << 1$ of strong axial confinement, 
which corresponds to the 2D regime, 
the 2D NPSE becomes a 2D nonlinear Schr\"odinger equation 
with cubic nonlinearity, also called 2D GPE \cite{7}. 
\par 
In the case $W(x,y)=0$, which corresponds to a 
homogeneous BEC in the cylindrical radial direction 
with periodic boundary conditions, the chemical potential reads 
\beq 
\mu = \gamma {\rho_2 \over \eta } + {1\over 4} 
\left({1 \over \eta^2 } + \eta^2 \right) 
= {5\over 4} \eta^2 - {3\over 4\eta^2} \; .  
\eeq  
It is easy to show that Eq. (4) reduces to 
$\mu = (5/4) (g\rho)^{1/2}$ in the 3D regime 
($\eta = (\gamma \rho_2)^{1/3}$) and it becomes 
$\mu = g\rho_2 +1/2$ in the 1D regime ($\eta =1$). 
The term $1/2$ is the axial energy in scaled units. 
From the 2D NPSE one calculates the radial velocity $c_2$ 
of first sound as 
$ 
c_2 = ( \rho_2 {\partial \mu\over \partial \rho_2} )^{1/2}
$.  

\begin{figure}
\centerline{\psfig{file=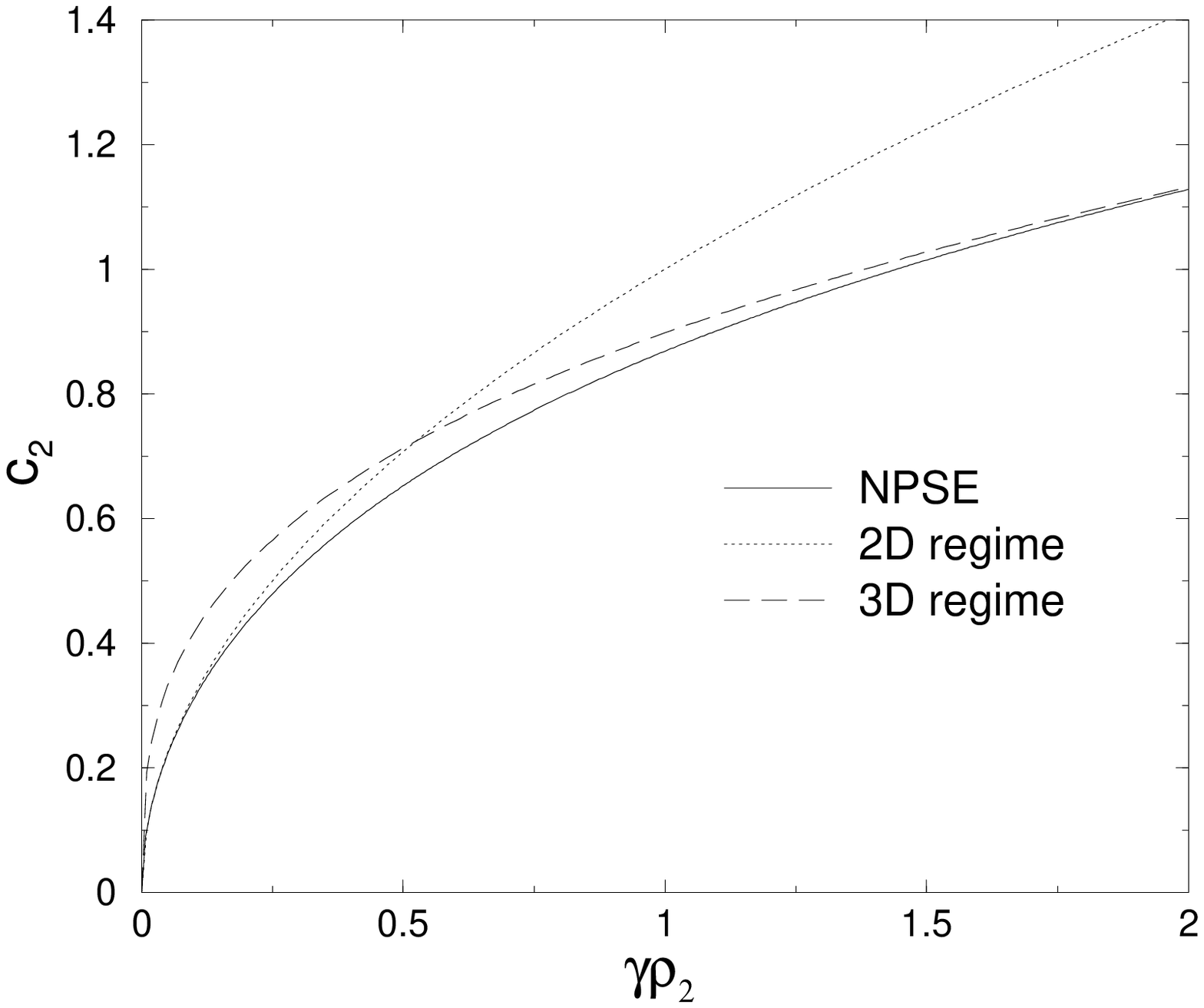,height=2.in}}
\small{FIG. 4: First sound radial velocity $c_2$ versus 
inter-atomic strength $\gamma\rho_2$, where 
$\gamma=2\sqrt{2\pi}a_s/a_z$ with $a_s$ scattering length and 
$a_z$ radial harmonic length. 
$\rho_2$ is the radial density of the disc-shaped 
radially-homogeneous BEC. 
Length in units $a_z$, density in units $1/a_z^2$ and 
time in units $1/\omega_z$. } 
\end{figure}  

In Figure 4 we plot $c_2$ as a function of $\gamma \rho_2$. 
The velocity of sound is 
$c_2=(5/6)(\gamma\rho_2)^{1/3}$ in the 3D regime  
and it becomes $c_2 = \sqrt{\gamma \rho_2}$ in the 2D regime. 
Our 2D value for $c_2$ is in full agreement 
with the result obtained from 2D GPE 
in the TF approximation, 
while the 3D value of $c_2$ differs from the exact one, 
obtained from the 3D GPE, only for the constant that is 
$3^{2/3}\pi^{1/3}/2^{7/3}$ instead of $5/6$. 
This error (less than 2\%)  in the 3D regime is a 
consequence of using in the 2D NPSE 
a Gaussian axial wave function. 
Thus, our sound velocity 
$c_2$ interpolates remarkably well 
between the exact results of low density (2D regime) 
and high density (3D regime). 
\par 
In the case $W(x,y)=m\omega_r^2(x^2+y^2)/2$ 
the collective oscillations of the BEC 
can be obtained from Eq. (3) 
with the condition of disc-shaped BEC, 
i.e. $0<\lambda < 1$. 
In Figure 5 we plot monopole and quadrupole frequencies 
of the BEC with a weak anisotropy. 
The monopole frequency $\omega_M$ does not depend on $\lambda$ 
in the limit of very small inter-atomic 
strength and coincides with the ideal gas limiting value. 
In the regime of mean-field 
bosons, $\omega_M$ changes with $\lambda$ from 
$(\omega_M/\omega_z)^2=5$ (isotropic TF 3D) 
to $(\omega_M/\omega_z)^2=3$ (anisotropic TF 3D). 
In a similar way the quadrupole frequency $\omega_Q$ moves from 
$(\omega_Q/\omega_r)^2=2$ (isotropic TF 3D) 
to $(\omega_Q/\omega_r)^2=10/3$ (anisotropic TF 3D), 
as predicted by Stringari \cite{12}. 

\begin{figure}
\centerline{\psfig{file=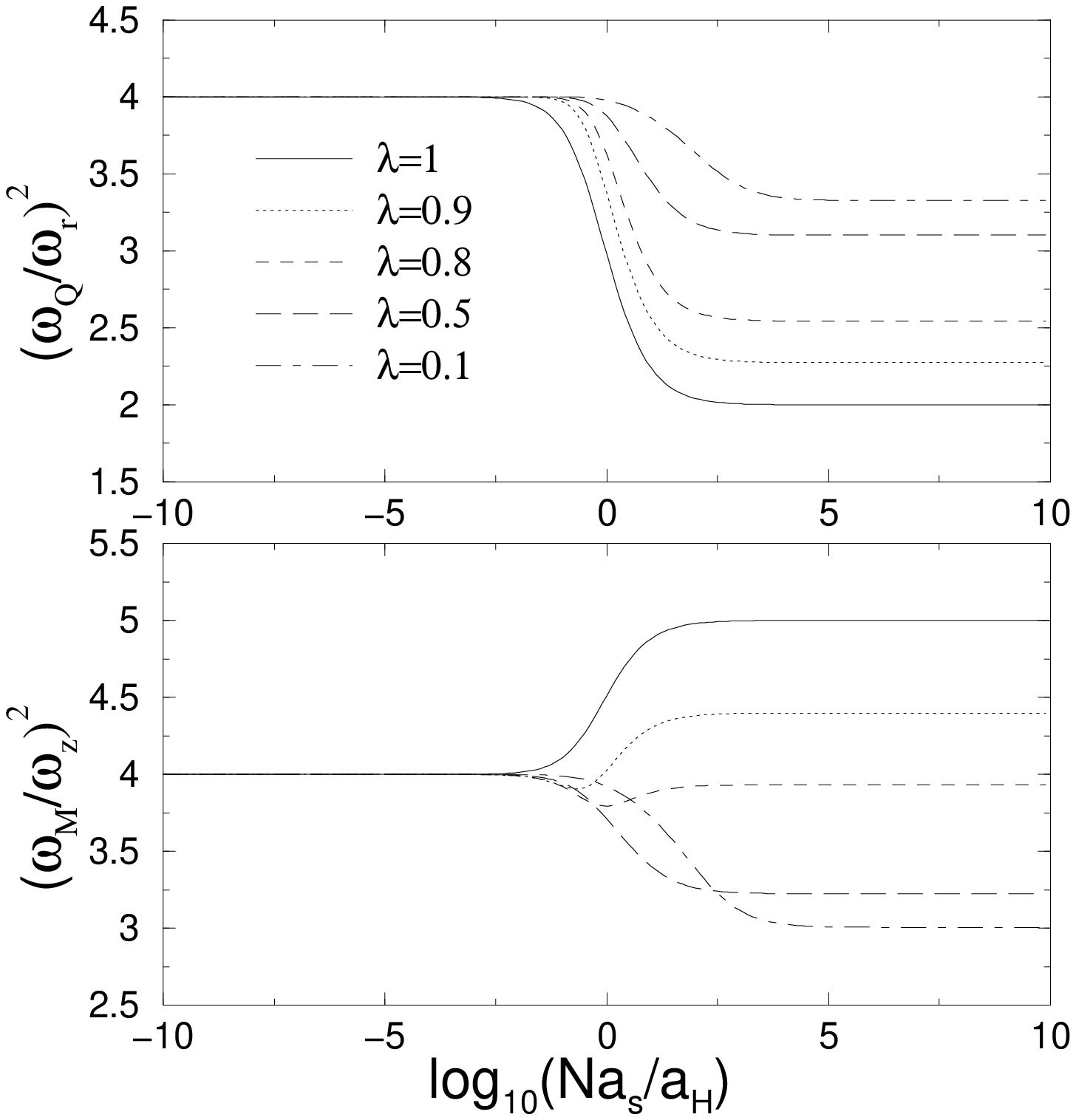,height=2.3in}}
\small{FIG. 5: Monopole frequency $\omega_M$ and quadrupole 
frequency $\omega_Q$ of collective oscillations 
for a disc-shaped BEC. 
Case of positive scattering length $a_s$ 
with small anisotropy.} 
\end{figure}  

By further increasing the disc-shaped anisotropy the main 
effect is that of shifting the transition from 
the ideal gas limit to the 3D regime at higher values of the 
interaction strength. 
As shown in Figure 6, both monopole and quadrupole frequencies 
display this behavior. Thus, these frequencies 
do not discriminate the regime of 2D confinement 
and mean-field bosons from the regime of 2D confinement 
and quasi-ideal bosons. 

\begin{figure}
\centerline{\psfig{file=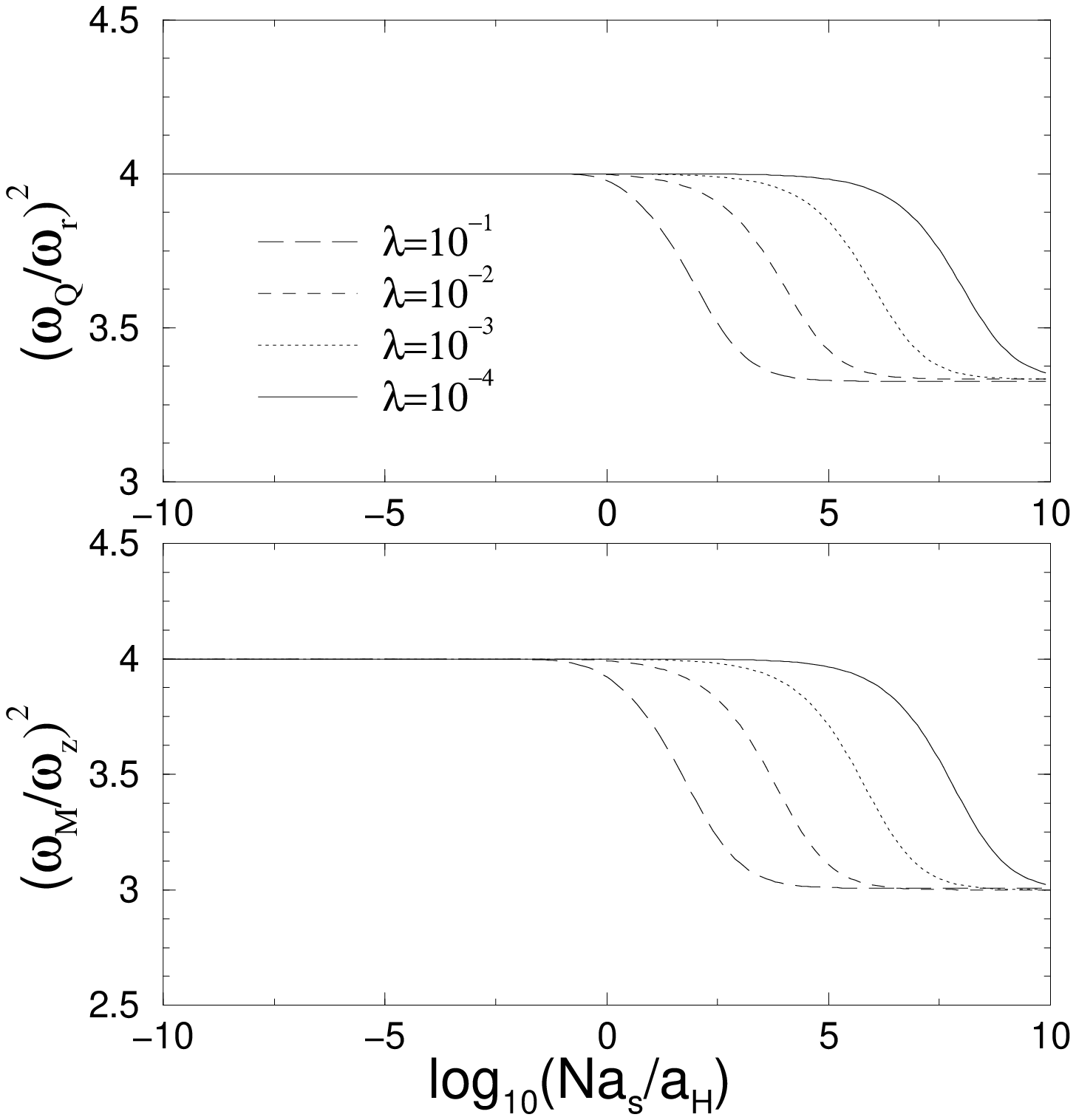,height=2.5in}}
\small{FIG. 6: Monopole frequency $\omega_M$ and quadrupole 
frequency $\omega_Q$ of collective oscillations 
for a disc-shaped BEC. 
Case of positive scattering length $a_s$ with 
large anisotropy.} 
\end{figure} 

The frequencies 
simply give a signature of the transition 
from ideal gas to TF 3D. 
Clearly, under strong confinement it is necessary 
to increase the inter-atomic interaction to 
mantain the 3D regime. Note that for the TF 2D regime
the dispersion relation of collective frequencies can 
be analytically obtained 
using the same hydrodynamical approach applied 
by Stringari for the TF 1D regime \cite{12}. 
In this way one recovers the result $\omega^2 = 4 \omega_r^2$ 
for the quadrupole frequency. 
\par 
The calculations with negative scattering 
length $a_s$ suggest that monopole and 
quadrupole oscillations of a highly deformed disc-shaped BEC 
do not depend on the inter-atomic strength. 

\begin{figure}
\centerline{\psfig{file=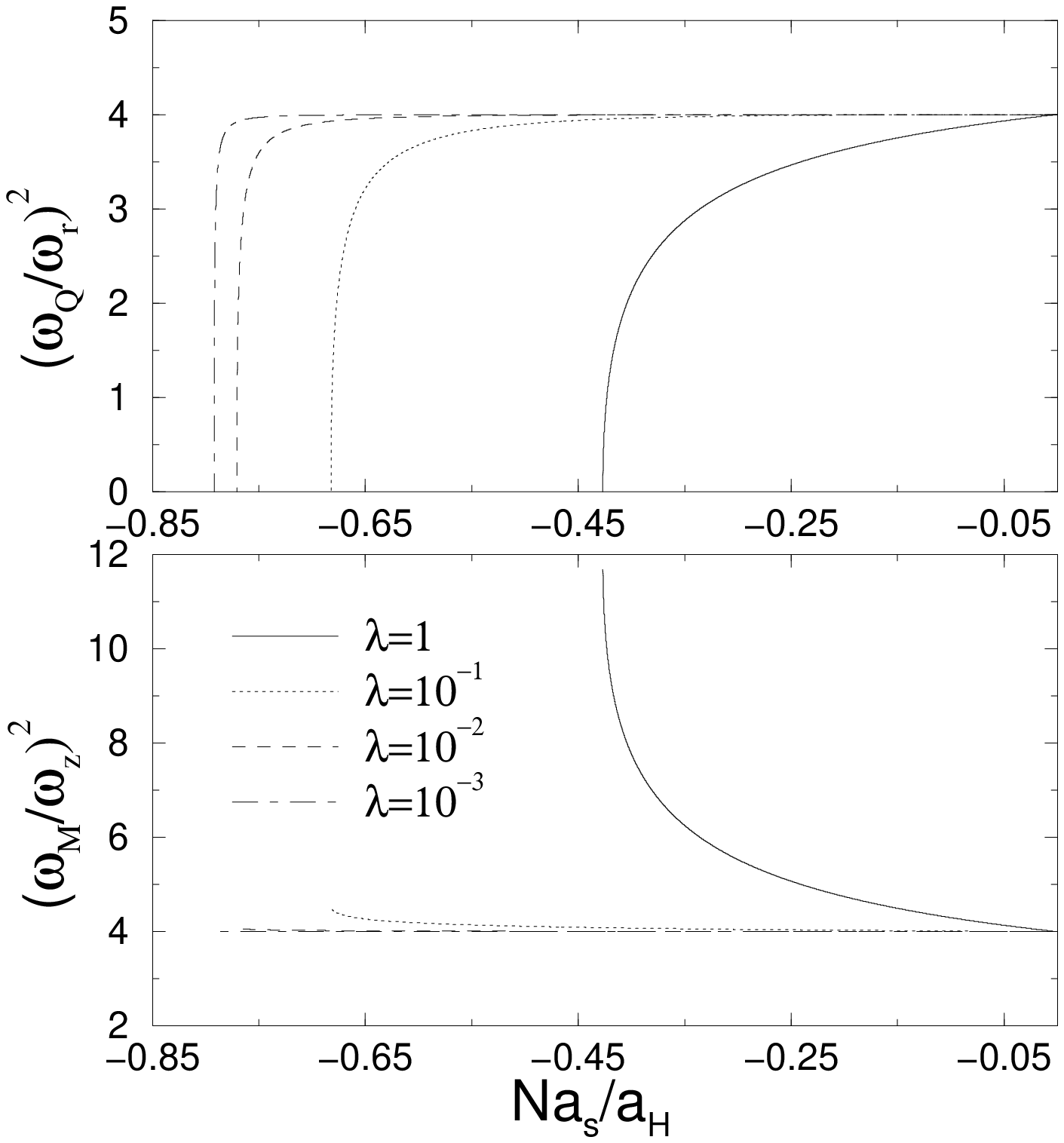,height=2.5in}}
\small{FIG. 7: Monopole frequency $\omega_M$ and quadrupole 
frequency $\omega_Q$ of collective oscillations 
for a disc-shaped BEC. 
Case of negative scattering length $a_s$.} 
\end{figure} 

As shown in Figure 7, only very close to the collapse $\omega_Q$  
goes to zero while $\omega_M$ remains 
essentially constant. 
\par 
In conclusion, our results exactly reproduces the ones obtained in specific 
dimensional regimes with the hydrodynamic method or 
the sum-rule method, but, 
they also give the crossover from one regime to another.

\end{document}